\documentclass[twocolumn,showpacs,preprintnumbers,amsmath,aps]{revtex4}
\usepackage{graphicx}
\usepackage{dcolumn}
\usepackage{bm}
\usepackage{multirow}
\usepackage{tabularx}
\usepackage{color}
\begin{document}
\def\eq#1{(\ref{#1})}
\def\fig#1{Fig.\hspace{1mm}\ref{#1}}
\def\tab#1{Tab.\hspace{1mm}\ref{#1}}
\title{
---------------------------------------------------------------------------------------------------------------\\
The thermodynamic properties of the high-pressure superconducting state in the hydrogen-rich compounds}
\author{R. Szcz{\c{e}}{\'s}niak, A.P. Durajski}
\email{adurajski@wip.pcz.pl}
\affiliation{Institute of Physics, Cz{\c{e}}stochowa University of Technology, Ave. Armii Krajowej 19, 42-200 Cz{\c{e}}stochowa, Poland}
\date{\today}
\begin{abstract}
The {\it ab initio} calculations suggest that the superconducting state in $\rm CaH_{6}$ under the pressure ($p$) at $150$ GPa has the highest critical temperature among the examined hydrogen-rich compounds. For this reason, the relevant thermodynamic parameters of the superconducting state in $\rm CaH_{6}$ have been determined; a wide range of the Coulomb pseudopotential has been assumed: $\mu^{\star}\in\left<0.1,0.3\right>$. 
It has been found that: 
(i) The critical temperature ($T_{C}$) changes in the range from $243$ K to $180$ K. 
(ii) The values of the ratio of the energy gap to the critical temperature ($R_{\Delta}\equiv 2\Delta\left(0\right)/k_{B}T_{C}$) can be found in the range from $5.42$ to $5.02$. 
(iii) The ratio of the specific heat jump ($\Delta C\left(T_{C}\right)$) to the value of the specific heat in the normal state ($C^{N}\left(T_{C}\right)$), which has been represented by the symbol $R_{C}$, takes the values from $3.30$ to $3.18$. 
(iv) The ratio $R_{H}\equiv T_{C}C^{N}\left(T_{C}\right)/H^{2}_{C}\left(0\right)$, where $H_{C}\left(0\right)$ denotes the critical thermodynamic field, changes from $0.122$ to $0.125$. 
The above results mean that even for the strong electron depairing correlations the superconducting state 
in $\rm CaH_{6}$ is characterized by a very high value of $T_{C}$, and the remaining thermodynamic parameters significantly deviate from the predictions of the BCS theory. 
The study has brought out the expressions that correctly predict the values of the thermodynamic parameters for the superconducting state in $\rm CaH_{6}$ and for the compounds: ${\rm SiH_{4}\left(H_{2}\right)_{2}}$, ${\rm Si_{2}H_{6}}$, ${\rm B_{2}H_{6}}$, ${\rm SiH_{4}}$, ${\rm GeH_{4}}$, and ${\rm PtH}$. 
Next, in the whole family of the hydrogen-rich compounds, the possible ranges of the values have been determined for $T_{C}$, $R_{\Delta}$, $R_{C}$, and $R_{H}$. 
It has been found that the maximum value of the critical temperature can be equal to $764$ K, which very well correlates with $T_{C}$ for metallic hydrogen ($p=2$ TPa). Other parameters ($R_{\Delta}$, $R_{C}$, and $R_{H}$) should not deviate from the predictions of the BCS theory more than the analogous parameters for $\rm CaH_{6}$.
\\
\\
Keywords: Superconductivity, Hydrogen-rich compounds, Thermodynamic properties, High-pressure effects
\end{abstract}
\pacs{74.20.Fg, 74.25.Bt, 74.62.Fj}
\maketitle

%
\section{INTRODUCTION}
%

The superconducting state in the hydrogen-rich compounds can be characterized by a very high value of the critical temperature \cite{Ashcroft}, \cite{Tse}, \cite{Gao}, \cite{Canales}, \cite{Gao1}, \cite{Chen}, \cite{Eremets}. It should be noted that when such situation takes place it is usually accompanied by a low value of the external pressure ($p$) in comparison to the pressure required for the metallization of hydrogen ($\sim 400$ GPa) \cite{Stadele}.

In the case of metallic hydrogen, for the pressures from $400$ GPa to $500$ GPa, the molecular phase exists, in which the superconducting state of the high critical temperature may be induced ($240$ K for $p=450$ GPa) \cite{Cudazzo01}, \cite{Cudazzo02}, \cite{Cudazzo03}, \cite{Zhang}. 
It should be noted that for certain values of the pressure (e.g. $p=414$ GPa), the superconducting state is strongly anisotropic \cite{Cudazzo01}, \cite{Cudazzo02}, \cite{Cudazzo03}. 

The numerical calculations carried out in the framework of the one- or multi-band model suggest that the thermodynamic properties of the superconducting state in the molecular hydrogen differ significantly from the expectations of the BCS theory \cite{Szczesniak1}, \cite{Szczesniak2}, \cite{Szczesniak3}, \cite{BCS1}, \cite{BCS2}. The above result is connected with the existence of the strong-coupling and retardation effects.

For higher pressures - in the range from $500$ GPa to $3.5$ TPa - the metallic phase of the atomic hydrogen is formed \cite{Yan}, \cite{Maksimov}, \cite{McMahon}, \cite{McMahon1}, \cite{Liu}. It has been found that the highest value of the critical temperature can be predicted for $p\simeq 2$ TPa, where $T_{C}$ is of the order of $600$-$700$ K \cite{Maksimov}, \cite{Szczesniak4}. In addition, the other thermodynamic parameters differ very significantly from the predictions of the BCS theory \cite{Szczesniak5}. In particular, the ratio of the energy gap to the critical temperature assumes values comparable to the values observed 
in the high-temperature superconductors (cuprates) \cite{Szczesniak4}, \cite{Szczesniak6}, \cite{Szczesniak7}.

In the case of the hydrogen-rich compounds the metalization can occur as early as in the interval of the pressures from $50$ GPa to $60$ GPa (e.g. $\rm SiH_{4}$) \cite{Chen}, \cite{Eremets}. Additionally, for $\rm SiH_{4}$ the existence of the superconducting state with the critical temperature of $17$ K has been found experimentally ($p=96$ GPa and $120$ GPa) \cite{Eremets}. Most probably, much higher values of the critical temperature can be observed in the compounds of the type:  ${\rm SiH_{4}\left(H_{2}\right)_{2}}$, ${\rm Si_{2}H_{6}}$, ${\rm B_{2}H_{6}}$, ${\rm GeH_{4}}$, and ${\rm PtH}$ \cite{Gao}, \cite{Li}, \cite{Jin}, \cite{Kazutaka}, \cite{Kim}.

It should be noted that in the pure elements under the influence of the high pressure the superconducting state with the relatively high critical temperature is also induced at this point the experimental results obtained for lithium and calcium are worth mentioning. In the case of lithium, the maximum value of the critical temperature is equal to $14$ K for $p=30.2$ GPa \cite{Deemyad}; whereas calcium is characterized by the superconducting state of the highest critical temperature among the pure elements ($T_{C}=25$ K for $161$ GPa) \cite{Yabuuchi}. Additionally, in 2011, it was suggested that 
$\left[T_{C}\right]_{\rm Ca}$ can be equal to $29$ K ($p=216$ GPa) \cite{Sakata}. However, this result has been challenged in the paper \cite{Andersson}. 

Let us note that the thermodynamic properties of the superconducting state of lithium and calcium differ very significantly from the expectations of the BCS theory \cite{Szczesniak8}, \cite{Szczesniak9}, \cite{Szczesniak10}, \cite{Szczesniak11}, \cite{Szczesniak12}, \cite{Szczesniak13}. It is also important to describe their properties, it is necessary to assume anomalously high values of the Coulomb pseudopotential ($\mu^{\star}$). This proves the existence of the strong electron depairing correlations in the examined elements 
($\left[\mu^{\star}\right]^{\rm \left(Li\right)}_{p=22.3 {\rm GPa}}=0.22$, $\left[\mu^{\star}\right]^{\rm \left(Li\right)}_{p=29.7 {\rm GPa}}=0.36$, and $\left[\mu^{\star}\right]^{\rm \left(Ca\right)}_{p=161 {\rm GPa}}=0.24$) \cite{Szczesniak8}, \cite{Szczesniak13}.

Recently, the branch literature has suggested that the critical temperature of the superconducting state can reach a very high value in the $\rm CaH_{6}$ compound ($p=150$ GPa) \cite{Wang}. As a result of the conducted analysis, the authors of the work \cite{Wang} have found that $T_{C}$ can be found in the range from $235$ K to $220$ K for the Coulomb pseudopotential from $0.1$ to $0.13$. Such unusually high values of the critical temperature suggest the existence of the superconducting state with strongly anomalous thermodynamic properties. However, it is unclear whether the value of the critical temperature for $\rm CaH_{6}$ will not to significantly decrease when too large values of $\mu^{\star}$ are assumed. This possibility cannot be excluded {\it a priori} taking into account the results obtained for the superconducting state in calcium \cite{Szczesniak9}, \cite{Szczesniak13}. 

Due to extremely interesting results obtained in the publication \cite{Wang}, we have determined the values of all relevant thermodynamic parameters characterizing the superconducting state in $\rm CaH_{6}$. We have taken into consideration a wide range of the Coulomb pseudopotential: $\mu^{\star}\in\left<0.1,0.3\right>$. The results have been then generalized, so as to be able to characterize the superconducting state in the entire group of the hydrogen-rich compounds.     
 
%
\section{THE FORMALISM}
%

The thermodynamic properties of the superconducting state in $\rm CaH_{6}$ have been determined with the help of the Eliashberg equations in the mixed representation \cite{Marsiglio}: 

\begin{widetext}
\begin{eqnarray}
\label{r1}
\phi\left(\omega+i\delta\right)&=&
                                  \frac{\pi}{\beta}\sum_{m=-M}^{M}
                                  \left[\lambda\left(\omega-i\omega_{m}\right)-\mu^{\star}\theta\left(\omega_{c}-|\omega_{m}|\right)\right]
                                  \frac{\phi_{m}}
                                  {\sqrt{\omega_m^2Z^{2}_{m}+\phi^{2}_{m}}}\\ \nonumber
                              &+& i\pi\int_{0}^{+\infty}d\omega^{'}\alpha^{2}F\left(\omega^{'}\right)
                                  \left[\left[N\left(\omega^{'}\right)+f\left(\omega^{'}-\omega\right)\right]
                                  \frac{\phi\left(\omega-\omega^{'}+i\delta\right)}
                                  {\sqrt{\left(\omega-\omega^{'}\right)^{2}Z^{2}\left(\omega-\omega^{'}+i\delta\right)
                                  -\phi^{2}\left(\omega-\omega^{'}+i\delta\right)}}\right]\\ \nonumber
                              &+& i\pi\int_{0}^{+\infty}d\omega^{'}\alpha^{2}F\left(\omega^{'}\right)
                                  \left[\left[N\left(\omega^{'}\right)+f\left(\omega^{'}+\omega\right)\right]
                                  \frac{\phi\left(\omega+\omega^{'}+i\delta\right)}
                                  {\sqrt{\left(\omega+\omega^{'}\right)^{2}Z^{2}\left(\omega+\omega^{'}+i\delta\right)
                                  -\phi^{2}\left(\omega+\omega^{'}+i\delta\right)}}\right],
\end{eqnarray}
and
\begin{eqnarray}
\label{r2}
Z\left(\omega+i\delta\right)&=&
                                  1+\frac{i}{\omega}\frac{\pi}{\beta}\sum_{m=-M}^{M}
                                  \lambda\left(\omega-i\omega_{m}\right)
                                  \frac{\omega_{m}Z_{m}}
                                  {\sqrt{\omega_m^2Z^{2}_{m}+\phi^{2}_{m}}}\\ \nonumber
                              &+&\frac{i\pi}{\omega}\int_{0}^{+\infty}d\omega^{'}\alpha^{2}F\left(\omega^{'}\right)
                                  \left[\left[N\left(\omega^{'}\right)+f\left(\omega^{'}-\omega\right)\right]
                                  \frac{\left(\omega-\omega^{'}\right)Z\left(\omega-\omega^{'}+i\delta\right)}
                                  {\sqrt{\left(\omega-\omega^{'}\right)^{2}Z^{2}\left(\omega-\omega^{'}+i\delta\right)
                                  -\phi^{2}\left(\omega-\omega^{'}+i\delta\right)}}\right]\\ \nonumber
                              &+&\frac{i\pi}{\omega}\int_{0}^{+\infty}d\omega^{'}\alpha^{2}F\left(\omega^{'}\right)
                                  \left[\left[N\left(\omega^{'}\right)+f\left(\omega^{'}+\omega\right)\right]
                                  \frac{\left(\omega+\omega^{'}\right)Z\left(\omega+\omega^{'}+i\delta\right)}
                                  {\sqrt{\left(\omega+\omega^{'}\right)^{2}Z^{2}\left(\omega+\omega^{'}+i\delta\right)
                                  -\phi^{2}\left(\omega+\omega^{'}+i\delta\right)}}\right], 
\end{eqnarray}
\end{widetext}
%
where the symbols $\phi\left(\omega\right)$ ($\phi_{m}\equiv\phi\left(i\omega_{m}\right)$) and $Z\left(\omega\right)$ 
($Z_{m}\equiv Z\left(i\omega_{m}\right)$) denote the order parameter function and the wave function renormalization factor 
on the real (imaginary $i\equiv\sqrt{-1}$) axis, respectively. The Matsubara frequency is represented by the formula:
$\omega_{m}\equiv \left(\pi / \beta\right)\left(2m-1\right)$, where $\beta\equiv\left(k_{B}T\right)^{-1}$; $k_{B}$ is the Boltzmann constant. 
The order parameter is defined by the ratio: $\Delta\equiv \phi/Z$.

The pairing kernel for the electron-phonon interaction has the form:
$\lambda\left(z\right)\equiv 2\int_0^{\Omega_{\rm{max}}}d\Omega\frac{\Omega}{\Omega ^2-z^{2}}\alpha^{2}F\left(\Omega\right)$.
The Eliashberg function ($\alpha^{2}F\left(\Omega\right)$) for $\rm CaH_{6}$ has been determined in the paper \cite{Wang}. The value of the maximum phonon frequency ($\Omega_{\rm max}$) is equal to $244.7$ meV. 

Let us notice that knowing the Eliashberg function allows us to calculate the electron-phonon coupling constant $\lambda\equiv 2\int^{\Omega_{\rm{max}}}_{0}d\Omega\frac{\alpha^{2}F\left(\Omega\right)}{\Omega}$. In the case of $\rm CaH_{6}$, we have the value $\lambda=2.69$. From the physical point of view, the above result means that in $\rm CaH_{6}$ a very strong coupling exists between the electrons and the crystal lattice vibrations.

The quantity $\theta$ in Eq. \eq{r1} denotes the Heaviside unit function; $\omega_{c}$ is the cut-off frequency ($\omega_{c}=3\Omega_{\rm{max}}$).

The symbols $N\left(\omega\right)$ and $f\left(\omega\right)$ represent the functions of Bose-Einstein and Fermi-Dirac, respectively.

The order parameter function and the wave function renormalization factor on the imaginary axis have been calculated using the equations \cite{Eliashberg1}, \cite{Eliashberg2}, \cite{Eliashberg3}, \cite{Eliashberg4}:
\begin{equation}
\label{r3}
\phi_{n}=\frac{\pi}{\beta}\sum_{m=-M}^{M}
\frac{\lambda\left(i\omega_{n}-i\omega_{m}\right)-\mu^{\star}\theta\left(\omega_{c}-|\omega_{m}|\right)}
{\sqrt{\omega_m^2Z^{2}_{m}+\phi^{2}_{m}}}\phi_{m},
\end{equation}
\begin{equation}
\label{r4}
Z_{n}=1+\frac{1}{\omega_{n}}\frac{\pi}{\beta}\sum_{m=-M}^{M}
\frac{\lambda\left(i\omega_{n}-i\omega_{m}\right)}{\sqrt{\omega_m^2Z^{2}_{m}+\phi^{2}_{m}}}
\omega_{m}Z_{m},
\end{equation}
where $M=1100$.

We can notice that the Eliashberg equations have been numerically solved with the use of the methods presented in the papers: \cite{Szczesniak14}, \cite{Szczesniak15}, \cite{Szczesniak16}, \cite{Szczesniak17}. The stable solutions have been obtained in the temperature range from $T_{0}=10$ K to $T_{C}$.

%
\section{THE CRITICAL TEMPERATURE FOR $\rm CaH_{6}$}
%

\begin{figure}[ht]
\includegraphics*[width=\columnwidth]{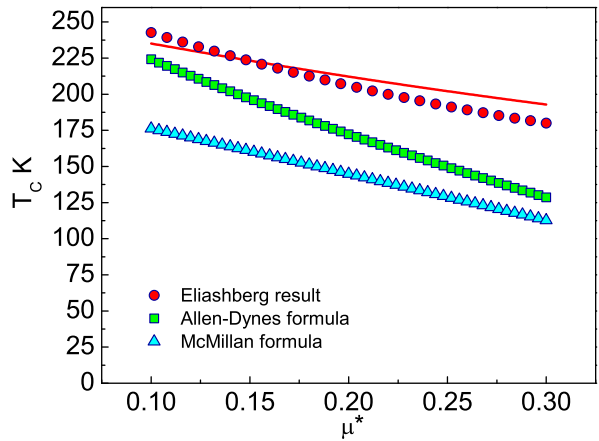}
\caption{The effect of the influence of the Coulomb pseudopotential on the critical temperature. The circles represent the results obtained from the Eliashberg equations. The solid line has been plotted using the Eq. \eq{r5}. The squared and triangle courses have been obtained using the classical Allen-Dynes and McMillan formula \cite{AllenDynes}, \cite{McMillan}.} 
\label{f1}
\end{figure}

In the first step, we have determined the dependence of the critical temperature on the Coulomb pseudopotential 
($\mu^{\star}\in\left<0.1,0.3\right>$). It has been found that the value of the critical temperature varies in the range from $243$ K to $180$ K 
(see \fig{f1}). The obtained result means that even in the case of the strong electron depairing correlations the value of the critical temperature in $\rm CaH_{6}$ is very high.

%
%

In addition, the classical analytical formulas (the Allen-Dynes or McMillan expresion \cite{AllenDynes}, \cite{McMillan}) significantly underestimate the critical temperature, especially for the high values of the Coulomb pseudopotential (see also \fig{f1}). A relatively good agreement between the analytical and numerical results can be obtained only for $\mu^{\star}\sim 0.1$ using the Allen-Dynes approach.

Due to the problem mentioned above, a new expression of the critical temperature has been tested. The examined formula has been recently postulated by ourselves for the ${\rm SiH_{4}\left(H_{2}\right)_{2}}$ compound ($p=250$ GPa) \cite{Szczesniak18}: 
\begin{equation}
\label{r5}
k_{B}T_{C}=f_{1}f_{2}\frac{\omega_{{\rm ln}}}{1.37}\exp\left[\frac{-1.125\left(1+\lambda\right)}{\lambda-\mu^{\star}}\right],
\end{equation}
where the functions $f_{1}$ and $f_{2}$ are defined by \cite{AllenDynes}:
$f_{1}\equiv\left[1+\left(\frac{\lambda}{\Lambda_{1}}\right)^{\frac{3}{2}}\right]^{\frac{1}{3}}$ and  
$f_{2}\equiv 1+\frac{\left(\frac{\sqrt{\omega_{2}}}{\omega_{\rm{ln}}}-1\right)\lambda^{2}}{\lambda^{2}+\Lambda^{2}_{2}}$.
The second moment of the normalized weight function ($\omega_{2}$) and the logarithmic frequency ($\omega_{{\rm ln}}$) can be calculated using the following expressions: 
$\omega_{2}\equiv\frac{2}{\lambda}\int^{\Omega_{\rm{max}}}_{0}d\Omega\alpha^{2}F\left(\Omega\right)\Omega$ and 
$\omega_{{\rm ln}}\equiv \exp\left[\frac{2}{\lambda}\int^{\Omega_{\rm{max}}}_{0}d\Omega\frac{\alpha^{2}F\left(\Omega\right)}
{\Omega}\ln\left(\Omega\right)\right]$. 

In the case of $\rm CaH_{6}$, we have achieved the following results: $\sqrt{\omega_{2}}=104$ $\rm{meV}$ and $\omega_{{\rm ln}}=88.8$ meV. 

The functions $\Lambda_{1}$ and $\Lambda_{2}$ have the form:
$\Lambda_{1}=2-0.14\mu^{\star}$, $\Lambda_{2}=\left(0.27+10\mu^{\star}\right)\left(\sqrt{\omega_{2}}/\omega_{\ln}\right)$.

On the basis of the results presented in \fig{f1}, it can be easily noticed that the modified Allen-Dynes formula properly reproduces the numerical results. 

Referring to the results included in the paper \cite{Wang}, it has been stated that the values of the critical temperature presented there are undervalued in comparison to the values of $T_{C}$ obtained with the presented method. For instance, in the work  \cite{Wang} the following has been obtained: 
$T_{C}\left(\mu^{\star}=0.1\right)=235$ K and $T_{C}\left(\mu^{\star}=0.3\right)=152$ K, while the result of our approach are:
$T_{C}\left(\mu^{\star}=0.1\right)=243$ K and $T_{C}\left(\mu^{\star}=0.3\right)=180$ K. The indicated differences in the predictions of the critical temperature probably result from the approximations used for the determination of $T_{C}$ in the paper \cite{Wang}.

%
\section{THE RANGE OF THE CRITICAL TEMPERATURE IN THE HYDROGEN-RICH COMPOUNDS}
%

In the paragraph, we have estimated the maximum value of the critical temperature for the hydrogen-rich compounds, which can be obtained assuming the reasonable value of the coupling constant and the Coulomb pseudopotential. In particular: 
$\lambda\in\left<1,3\right>$ and $\mu^{\star}\in\left<0.1,0.3\right>$.

\begin{figure}[ht]
\includegraphics*[width=\columnwidth]{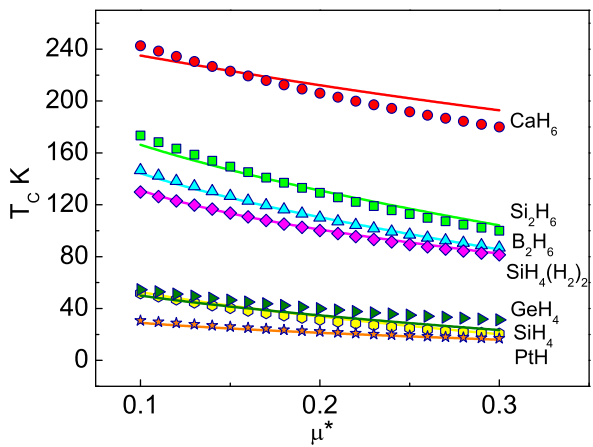}
\caption{The critical temperature for the selected hydrogen-rich compounds as a function of the Coulomb pseudopotential. The symbols represent the strict results obtained with the help of the Eliashberg equations: 
for ${\rm Si_{2}H_{6}}$ ($p=275$ GPa) from the paper \cite{Szczesniak19}, 
for ${\rm B_{2}H_{6}}$ ($p=360$ GPa) from \cite{Szczesniak20},  
for ${\rm SiH_{4}\left(H_{2}\right)_{2}}$ ($p=250$ GPa) from \cite{Szczesniak18}, \cite{Durajski}, 
for ${\rm GeH_{4}}$ ($p=20$ GPa) from \cite{Szczesniak21},
for ${\rm SiH_{4}}$ ($p=250$ GPa) from \cite{Szczesniak22}, 
and for ${\rm PtH}$ ($p=76$ GPa) from \cite{Szczesniak23}. 
The lines have been obtained using the formula \eq{r5}.} 
\label{f2}
\end{figure}

The critical temperature has been calculated using the formula \eq{r5}. The first step was to verify whether under consideration one can obtain sufficiently accurate $T_{C}$ values compared to the values determined with the use of the Eliashberg equations. Beside ${\rm CaH_{6}}$, the hydrogenated compounds characterized by a very high critical temperature have been chosen: 
${\rm SiH_{4}\left(H_{2}\right)_{2}}$, ${\rm Si_{2}H_{6}}$, ${\rm B_{2}H_{6}}$, ${\rm SiH_{4}}$, ${\rm GeH_{4}}$ and ${\rm PtH}$ \cite{Li}, \cite{Jin}, \cite{Kazutaka}, \cite{Gao}, \cite{Kim}. The results have been shown in \fig{f2}. It is easy to notice that the formula \eq{r5} reproduces results obtained by solving the Eliashberg equations with a very good accuracy.

At this point it should be again clearly underlined that the classical Allen-Dynes or McMillan formula should not be used, because the mentioned formulas significantly lower $T_{C}$.

%
%

In the simplest case, the Eliashberg function for the hydrogenated compounds can be modeled with the help of the following expression:
\begin{equation}
\label{r6}
\alpha^{2}F\left(\Omega\right)=\Omega_{1}\delta\left(\Omega-\Omega_{2}\right), 
\end{equation}
where the symbol $\delta\left(z\right)$ denotes the Dirac delta distribution. The parameters in Eq. \eq{r5} take the form: $f_{2}=1$, $\lambda=2\Omega_{1}/\Omega_{2}$, $\omega_{{\rm ln}}=\Omega_{2}$, and $\omega_{2}=\Omega^{2}_{2}$. Considering the above results, the expression for the critical temperature can be rewritten as:
\begin{eqnarray}
\label{r7}
k_{B}T_{C}=\frac{\Omega_{2}}{1.37}\left[1+\left(\frac{2\Omega_{1}}{\Lambda_{1}\Omega_{2}}\right)^{3/2}\right]^{1/3}\exp\left[\frac{1.125\left(\Omega_{2}+2\Omega_{1}\right)}{\mu^{\star}\Omega_{2}-2\Omega_{1}}\right]. 
\end{eqnarray}

\begin{figure}[ht]
\includegraphics*[width=\columnwidth]{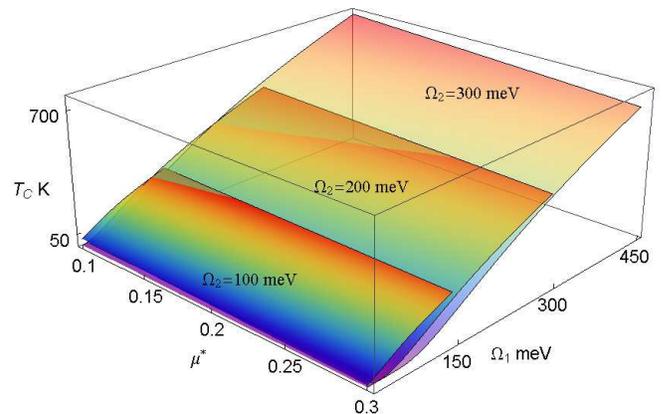}
\caption{The critical temperature as a function of $\mu^{\star}$ and $\Omega_{1}$ for the selected values of $\Omega_{2}$. The ranges of the values $\Omega_{1}$ have been selected in such a way that for specific $\Omega_{2}$ the maximum value of the electron-phonon coupling constant equals $3$.} 
\label{f3}
\end{figure}

The formula \eq{r7} has been used to determine the possible values of the critical temperature for $\lambda\in\left<1,3\right>$ and $\mu^{\star}\in\left<0.1,0.3\right>$. The results have been shown in \fig{f3}.

%
%

The obtained results suggest that for the reasonable values of the input parameters, the maximum critical temperature in the hydrogenated compounds may be equal to $764$ K. This result coincides with the estimation of the maximum critical temperature for the metallic atomic hydrogen ($p=2$ TPa) \cite{Szczesniak4}. Bearing in mind the fact that the hydrogen-rich compounds are used for the chemical pre-compression of hydrogen, there is a real chance of obtaining the superconducting state at the room temperature and at the pressure, which is much lower when compared to the pressure required for metallic hydrogen.

Finally, we have found that the formula \eq{r5} works very-well also for pure metallic hydrogen. We have illustrated this fact in the Appendix A.    

%
\section{THE ORDER PARAMETER FOR $\rm CaH_{6}$}
%

%
\begin{figure*}[ht]
\includegraphics*[scale=0.6]{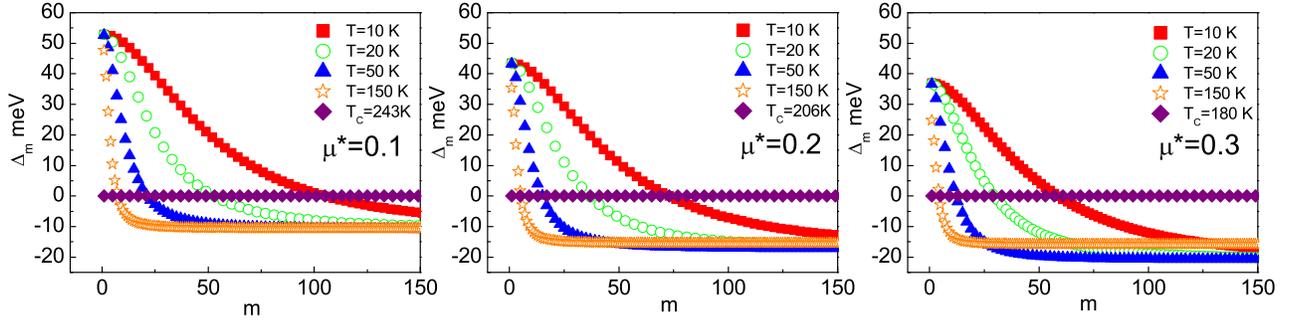}
\caption{The order parameter on the imaginary axis for the selected values of the temperature and the Coulomb pseudopotential.} 
\label{f4}
\end{figure*}
%

In \fig{f4}, the form of the order parameter on the imaginary axis for $\rm CaH_{6}$ compound has been presented. The selected values of the temperature and the Coulomb pseudopotential have been taken into account. It can be observed that with the increase of the $m$ parameter, the values of the function $\Delta_{m}$ are yielding to a strong decrease and then become saturated. Based on the presented data, it has been found that the increase in the Coulomb pseudopotential also causes the decrease of the order parameter.

\begin{figure}[ht]
\includegraphics*[width=\columnwidth]{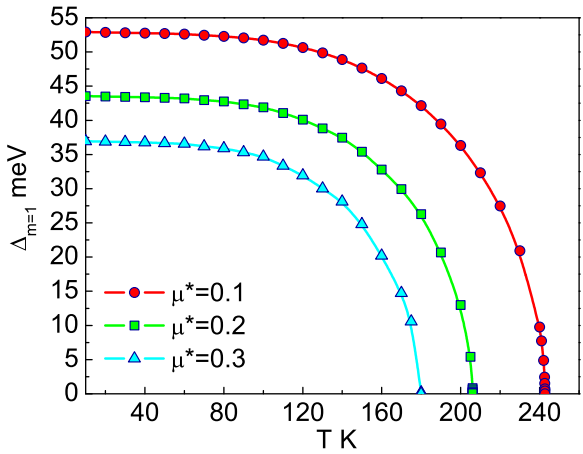}
\caption{The order parameter for the first Matsubara frequency as a function of the temperature for the selected values of the Coulomb pseudopotential. The functions presented in the figure can be parameterized by using the expression: 
$\Delta_{m=1}\left(T,\mu^{\star}\right)=\Delta_{m=1}\left(T_{0},\mu^{\star}\right)\sqrt{1-\left(\frac{T}{T_{C}}\right)^{\beta}}$, where: $\Delta_{m=1}\left(T_{0},\mu^{\star}\right)=138.9\left(\mu^{\star}\right)^{2}-135.7\mu^{\star}+65.2$ meV and $\beta=3.4$.} 
\label{f5}
\end{figure}

The temperature dependence on the order parameter can be represented by plotting the form of the function $\Delta_{m=1}\left(T\right)$. Note, that from the physical point of view, the quantity $2\Delta_{m=1}\left(T\right)$ reproduces the value of the energy gap at the Fermi level with the good approximation. The obtained results have been presented in \fig{f5}.

%
%

With the order parameter illustrated on the imaginary axis, it is possible (with the help of Eqs. \eq{r1} and \eq{r2}) to obtain the form of the order parameter on the real axis ($\Delta_{m}\rightarrow\Delta\left(\omega\right)$). Let us notice that the function $\Delta\left(\omega\right)$ unlike $\Delta_{m}$ takes complex values. In particular, its real part is used to calculate the physical value of the energy gap at the Fermi level, while the imaginary part determines the damping effects \cite{Varelogiannis}.

\begin{figure*}[ht]
\includegraphics*[scale=0.6]{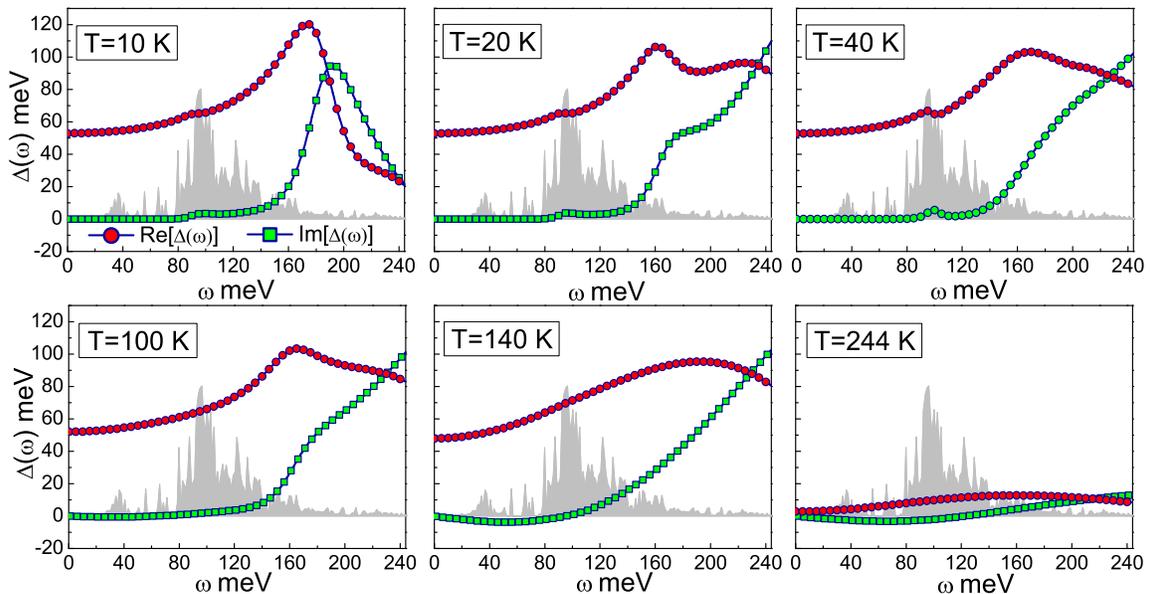}
\caption{The real part and the imaginary part of the order parameter on the real axis for the selected values of the temperature and $\mu^{\star}=0.1$. In addition, the figure presents the rescaled Eliashberg function ($20\alpha^{2}F\left(\Omega\right)$).} 
\label{f6}
\end{figure*}

The results obtained for $\mu^{\star}=0.1$ have been presented in \fig{f6}. It has been found that for the low frequency, the non-zero values are assumed only by the real part of $\Delta\left(\omega\right)$. 

Analyzing the data presented in \fig{f6}, it can be observed that in the range of low temperatures, the dependence of the order parameter on frequency has a more complicated character than for higher temperatures. Of particular note are the strong local maxima of the function 
${\rm Re}\left[\Delta\left(\omega\right)\right]$ and ${\rm Im}\left[\Delta\left(\omega\right)\right]$. This behavior is related to the fact that for low temperatures, the course of $\Delta\left(\omega\right)$ is strongly correlated with the shape of the Eliashberg function \cite{Szczesniak2}, \cite{Varelogiannis}.

\begin{figure*}[ht]
\includegraphics*[scale=0.6]{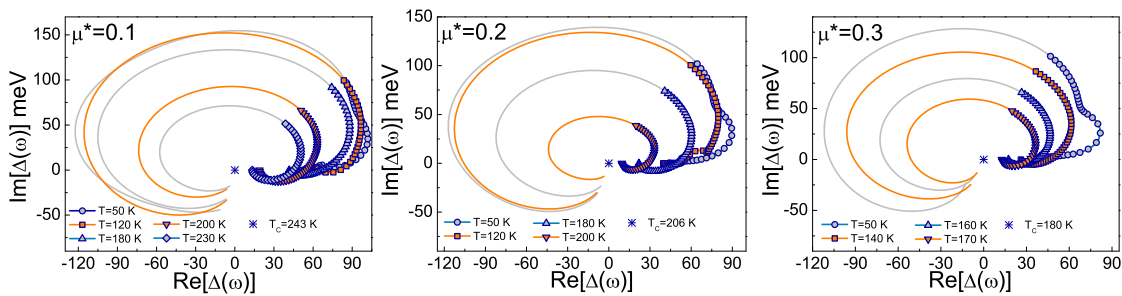}
\caption{The order parameter on the complex plane for the selected values of the temperature and the Coulomb pseudopotential. The lines with the symbols represent the values of $\Delta\left(\omega\right)$ for the frequency range from $0$ to $\Omega_{\rm max}$. The smooth lines correspond to the frequency range from $\Omega_{\rm max}$ to $\omega_{c}$.} 
\label{f7}
\end{figure*}

The form of the function $\Delta\left(\omega\right)$ for the selected values of the temperature and the Coulomb pseudopotential has been also plotted on the complex plane (see \fig{f7}). 

It can be noticed that the values of the order parameter form characteristic spirals of the radius decreasing together with the increasing temperature. Based on the presented results, the value of the frequency has been specified ($\omega_{a}$) for which 
the effective electron-electron interaction becomes depairing (${\rm Re}\left[\Delta\left(\omega\right)\right]\leq 0$) \cite{Varelogiannis}. In the case $\mu^{\star}=0.1$ we obtain: $\omega_{a}=320$ meV. In addition, it has been found that the increase of the Coulomb pseudopotential causes a significant decrease of $\omega_{a}$. In particular, for $\mu^{\star}=0.3$ we obtain: $\omega_{a}=289$ meV.

%
%

%
%
 
The physical value of the order parameter has been calculated using the formula below:

\begin{equation}
\label{r8}
\Delta\left(T\right)={\rm Re}\left[\Delta\left(\omega=\Delta\left(T\right)\right)\right].
\end{equation}

In particular, for $T=T_{0}$ it has been obtained: $\Delta\left(T_{0}\right)\equiv\Delta(0)\in\left<56.6,39.0\right>$ meV, for the range of the values $\mu^{\star}$ from $0.1$ to $0.3$. On that basis, the dimensionless ratio $R_{\Delta}\equiv 2\Delta\left(0\right)/k_{B}T_{C}$ has been determined. 
The result has the form: $R_{\Delta}\in\left<5.42,5.02\right>$. 

It should be noted that the values of the parameter $R_{\Delta}$ very significantly exceed the value predicted by the BCS theory ($R_{\Delta}=3.53$) \cite{BCS1}, \cite{BCS2}. 

The above result is related to the existence of the strong-coupling and retardation effects in $\rm CaH_{6}$. In the simplest case, it can be characterized by the ratio: $k_{B}T_{C}/\omega_{\rm ln}$. In the case of the BCS theory $0$ value is obtained. For $\rm CaH_{6}$, the examined ratio varies in the range from $0.156$ to $0.098$ ($\mu^{\star}\in\left<0.1,0.3\right>$).

It is easy to notice that the values of the parameter $R_{\Delta}$ require a number of complicated and time-consuming calculations. For this reason, the appropriate formula has been given below. It allows us to estimate the parameter $R_{\Delta}$ with a good approximation, depending on the assumed value of $\mu^{\star}$:

\begin{equation}
\label{r9}
\frac{R_{\Delta}}{\left[R_{\Delta}\right]_{{\rm BCS}}}=1+3.2\left(\frac{k_{B}T_{C}}{a\omega_{{\rm ln}}}\right)^{2}
\ln\left(\frac{a\omega_{{\rm \ln}}}{k_{B}T_{C}}\right),
\end{equation}
where $a=0.517$. In the case of $\rm CaH_{6}$, the differences between the numerical and the analytical results do not exceed $2$\%.

%
\section{THE RANGE OF THE PARAMETER $R_{\Delta}$ IN THE HYDROGEN-RICH COMPOUNDS}
%

\begin{figure}[ht]
\includegraphics*[width=\columnwidth]{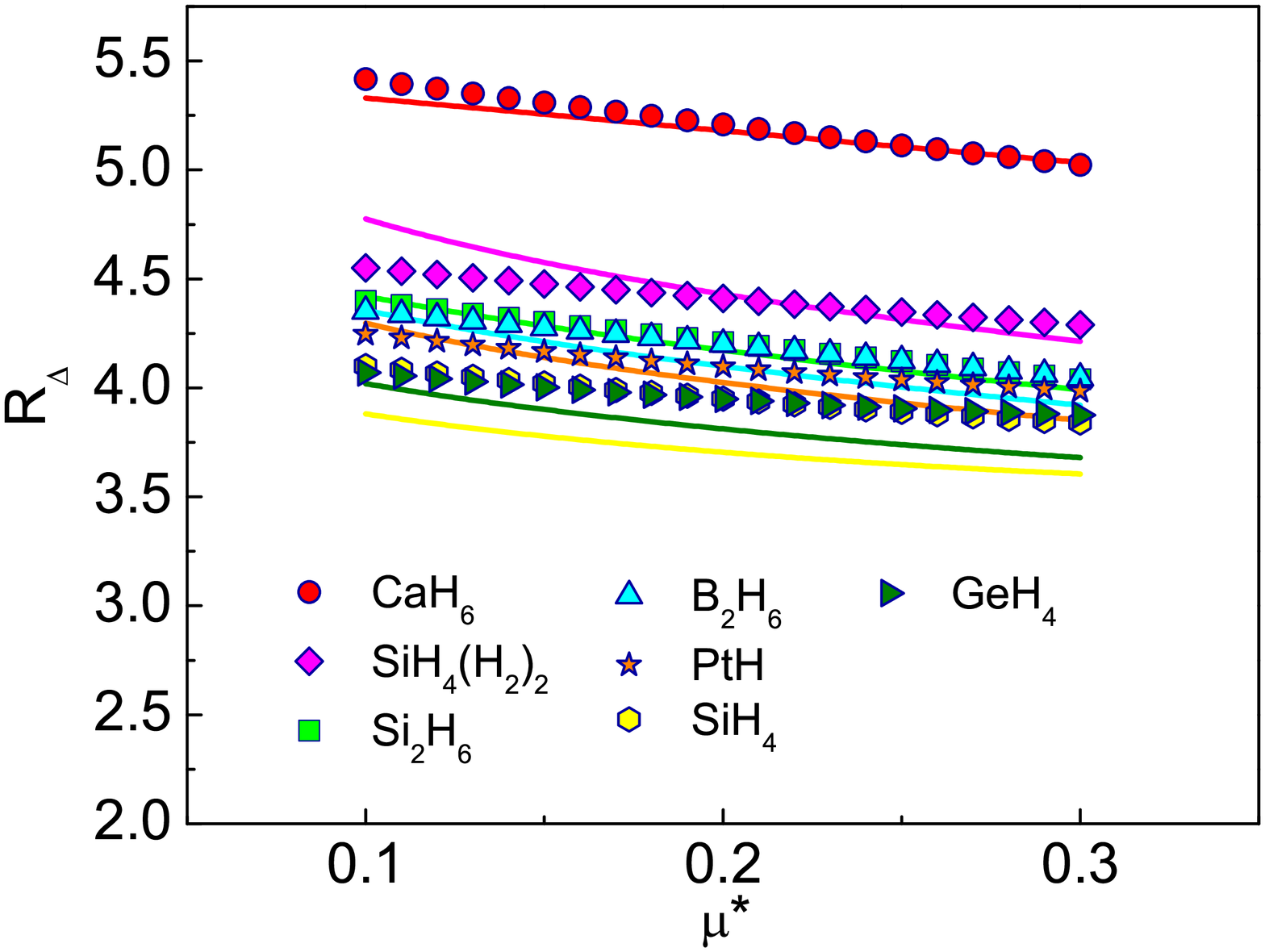}
\caption{The ratio $R_{\Delta}$ in the dependence of the assumed values of $\mu^{\star}$ for the selected hydrogen-rich compounds. The symbols represent the results obtained with the help of the Eliashberg equations: 
for ${\rm SiH_{4}\left(H_{2}\right)_{2}}$ ($p=250$ GPa) from the paper \cite{Szczesniak18}, \cite{Durajski},
for ${\rm Si_{2}H_{6}}$ ($p=275$ GPa) from \cite{Szczesniak19}, 
for ${\rm B_{2}H_{6}}$ ($p=360$ GPa) from \cite{Szczesniak20},  
for ${\rm PtH}$ ($p=76$ GPa) from \cite{Szczesniak23},
for ${\rm SiH_{4}}$ ($p=250$ GPa) from \cite{Szczesniak22}, and 
for ${\rm GeH_{4}}$ ($p=20$ GPa) from \cite{Szczesniak21}. 
The lines have been obtained using the formula \eq{r9}.}
\label{f8}
\end{figure}

The correctness of the formula \eq{r9} for $\rm CaH_{6}$ suggests its applicability for the other hydrogenated compounds. This hypothesis has been verified for those systems, for which accurate numerical data can be found in the literature. The results have been summarized in \fig{f8}.

%
%

Based on \fig{f8}, it can be easily seen that the expression \eq{r9} correctly predicts the dependence of $R_{\Delta}$ on $\mu^{\star}$ for the analyzed group of the systems. 

\begin{figure}[ht]
\includegraphics*[width=\columnwidth]{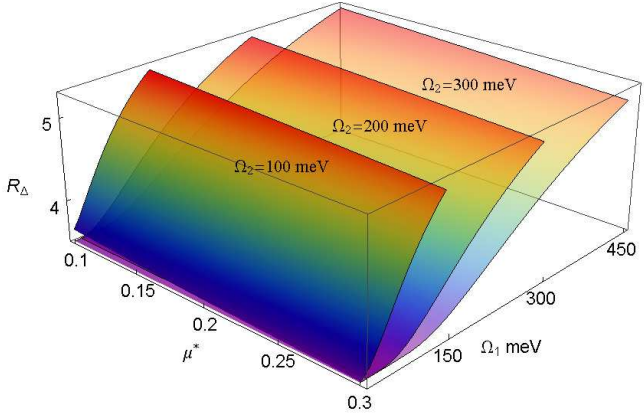}
\caption{The ratio $R_{\Delta}$ as a function of $\mu^{\star}$ and $\Omega_{1}$ for the selected values of $\Omega_{2}$. The ranges of the values $\Omega_{1}$ have been selected in such a way that for specific $\Omega_{2}$ the maximum value of the electron-phonon coupling constant equals $3$.} 
\label{f9}
\end{figure}

Generalizing the obtained result allowed the determination of the possible range of the $R_{\Delta}$ parameter in the entire group of the hydrogenated compounds. For this purpose the formula \eq{r9} and the Eliashberg function defined by the expression \eq{r6} have been used. The results have been summarized in \fig{f9}.

%
%

It has been found that the maximum value of the parameter $R_{\Delta}$ equals $5.27$. Taking into account the approximations which have been made, it has to be assumed that the value of the ratio $R_{\Delta}$ in the group of the hydrogenated compounds should not exceed the values obtained for $\rm CaH_{6}$.

%
\section{THE THERMODYNAMIC CRITICAL FIELD AND THE SPECIFIC HEAT FOR $\rm CaH_{6}$}
%

The thermodynamic critical field and the specific heat should be calculated using the formula for the free energy difference between the superconducting and the normal state \cite{Bardeen}: 

\begin{eqnarray}
\label{r10}
\frac{\Delta F}{\rho\left(0\right)}&=&-\frac{2\pi}{\beta}\sum_{n=1}^{M}
\left(\sqrt{\omega^{2}_{n}+\Delta^{2}_{n}}- \left|\omega_{n}\right|\right)\\ \nonumber
&\times&(Z^{S}_{n}-Z^{N}_{n}\frac{\left|\omega_{n}\right|}
{\sqrt{\omega^{2}_{n}+\Delta^{2}_{n}}}),  
\end{eqnarray}   
where $\rho\left(0\right)$ denotes the value of the density
of states at the Fermi level; $Z^{S}_{n}$ and $Z^{N}_{n}$ denote the wave function renormalization factors for the superconducting ($S$) and the normal ($N$) state, respectively. 

%
\begin{figure}[ht]
\includegraphics*[width=\columnwidth]{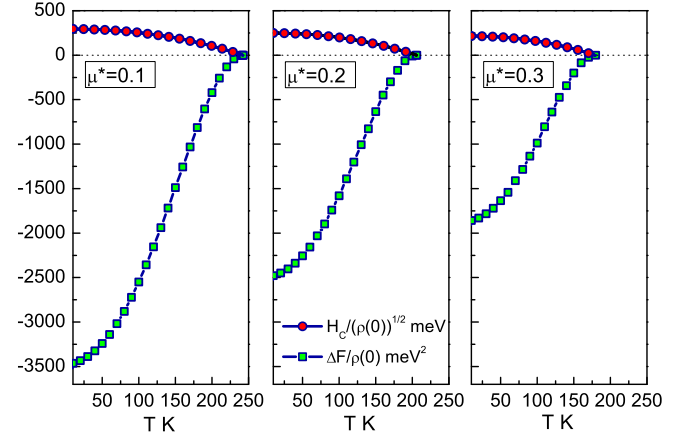}
\caption{The free energy and the thermodynamic critical field as a function of the temperature for the selected values of the Coulomb pseudopotential.} 
\label{f10}
\end{figure}
%

\fig{f10} presents the dependence of the ratio $\Delta F/\rho\left(0\right)$ on the temperature, where it is clear that with the increase of the Coulomb pseudopotential, the absolute value of $\Delta F$ strongly decreases. For instance: 
$\left[\Delta F\left(T_{0}\right)\right]_{\mu^{\star}=0.3}/\left[\Delta F\left(T_{0}\right)\right]_{\mu^{\star}=0.1}=0.54$.

On the basis of the expression \eq{r10}, the thermodynamic critical field has been determined:

\begin{equation}
\label{r11}
\frac{H_{C}}{\sqrt{\rho\left(0\right)}}=\sqrt{-8\pi\left[\Delta F/\rho\left(0\right)\right]}.
\end{equation}

Also \fig{f10} shows the obtained data. The destructive influence of the depairing electron correlations on the value of the thermodynamic critical field can be characterized by the ratio: $\left[H_{C}\left(0\right)\right]_{\mu^{\star}=0.3}/
\left[H_{C}\left(0\right)\right]_{\mu^{\star}=0.1}=0.73$, where it has been assumed that: $H_{C}\left(0\right)\equiv H_{C}\left(T_{0}\right)$.

The difference in the specific heat between the superconducting and the normal state ($\Delta C\equiv C^{S}-C^{N}$) should be calculated with the use of the following formula:

\begin{equation}
\label{r12}
\frac{\Delta C\left(T\right)}{k_{B}\rho\left(0\right)}=-\frac{1}{\beta}\frac{d^{2}\left[\Delta F/\rho\left(0\right)\right]}{d\left(k_{B}T\right)^{2}}.
\end{equation}

The specific heat of the normal state can be most conveniently estimated using the formula: 
$\frac{C^{N}\left(T\right)}{ k_{B}\rho\left(0\right)}=\frac{\gamma}{\beta}$, where the Sommerfeld constant is given by: $\gamma\equiv\frac{2}{3}\pi^{2}\left(1+\lambda\right)$. 

\begin{figure}[ht]
\includegraphics*[width=\columnwidth]{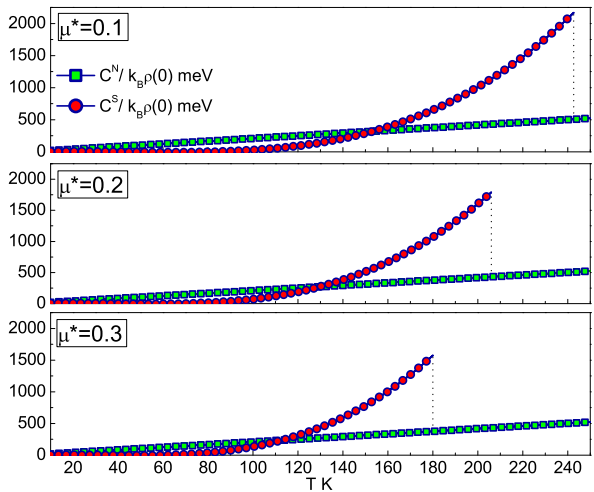}
\caption{The specific heat in the normal and superconducting state as a function of the temperature for the selected values of the Coulomb pseudopotential.} 
\label{f11}
\end{figure}

\fig{f11}, represents the dependence of the specific heat in the superconducting and the normal state as a function of the temperature for the selected values of the Coulomb pseudopotential. The characteristic jump at the critical temperature has been marked with the vertical line. The impact of the Coulomb pseudopotential on the value of the jump determines the ratio: 
$\left[\Delta C\left(T_{C}\right)\right]_{\mu^{\star}=0.3}/
\left[\Delta C\left(T_{C}\right)\right]_{\mu^{\star}=0.1}=0.72$.

%
%

Based on the results obtained for the thermodynamic critical field and the specific heat, the values of the two dimensionless parameters can be calculated:

\begin{equation}
\label{r13}
R_{H}\equiv\frac{T_{C}C^{N}\left(T_{C}\right)}{H_{C}^{2}\left(0\right)},
\qquad {\rm and} \qquad
R_{C}\equiv\frac{\Delta C\left(T_{C}\right)}{C^{N}\left(T_{C}\right)}.
\end{equation}

For the case of ${\rm CaH_{6}}$, the achieved results have been summarized in \tab{t1}. Let us note that ratios $R_{H}$ and $R_{C}$ in the classical BCS theory represent the universal constants of the model and reach the values of $0.168$ and $1.43$, respectively \cite{BCS1}, \cite{BCS2}. 
The results presented in \tab{t1} indicate that the properties of the superconducting state in ${\rm CaH_{6}}$ very significantly differ from the predictions of the BCS model.

%
\begin{table}
\caption{\label{t1} The dependence of the parameters $R_{H}$ and $R_{C}$ on the value of the Coulomb pseudopotential.}
\begin{ruledtabular}
\begin{tabular}{c|cc}
$\mu^{\star}$ & $R_{H}$ & $R_{C}$ \\
\hline
 $ $   & $ $     & $ $      \\
 $0.1$ & $0.122$ & $3.30$   \\                                                                                            
 $ $   & $ $     & $ $      \\
 $0.2$ & $0.123$ & $3.21$   \\                                                                                            
 $ $   & $ $     & $ $      \\
 $0.3$ & $0.125$ & $3.18$  \\\hline 
\end{tabular}
\end{ruledtabular}
\end{table}

%
\section{THE RANGE OF THE PARAMETERS $R_{H}$ AND $R_{C}$ IN THE HYDROGEN-RICH COMPOUNDS}
%

%
\begin{figure}[ht]
\includegraphics*[width=\columnwidth]{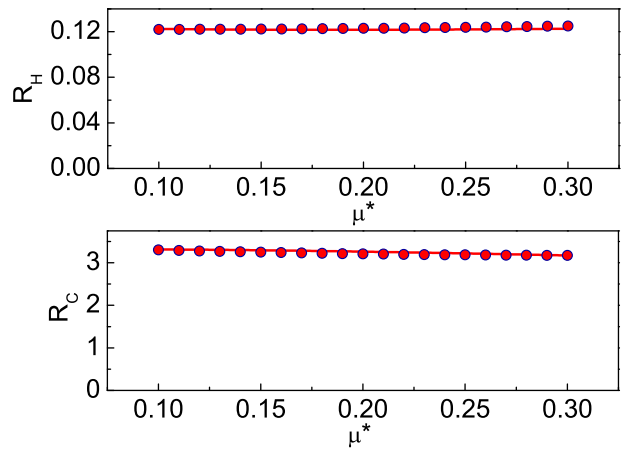}
\caption{The dependence of the parameters $R_{H}$ and $R_{C}$ on the value of the Coulomb pseudopotential for ${\rm CaH_{6}}$. The filled circles represent the numerical results. The lines have been obtained with the help of the expressions \eq{r14} and \eq{r15}.} 
\label{f12}
\end{figure}
%

%
\begin{figure*}[ht]
\includegraphics*[scale=0.25]{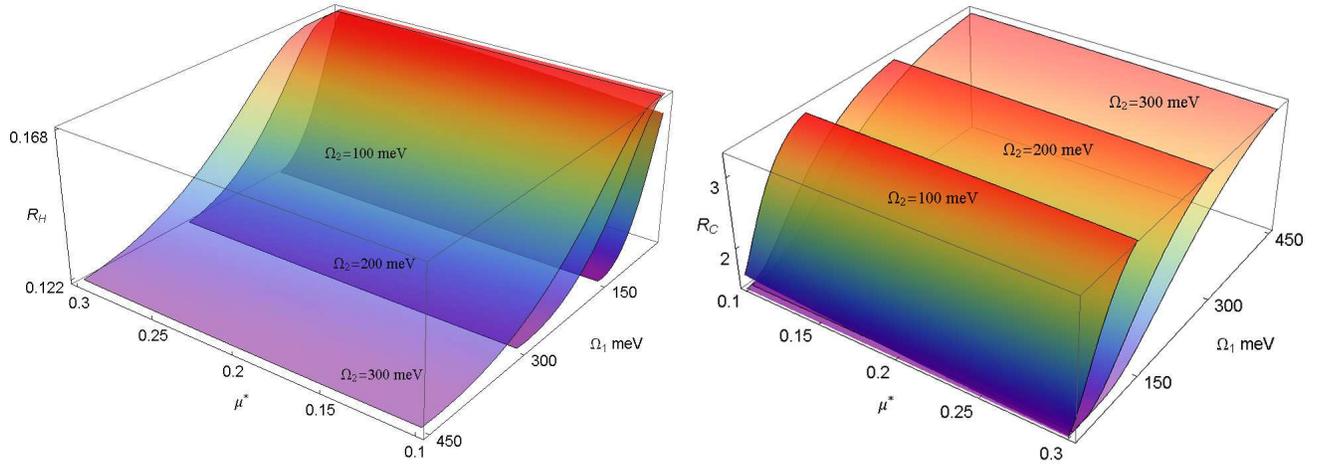}
\caption{The ratios $R_{H}$ and $R_{C}$ as functions of $\mu^{\star}$ and $\Omega_{1}$ for the selected values of $\Omega_{2}$. The intervals of the values $\Omega_{1}$ have been selected in such a way that for the assumed $\Omega_{2}$ the maximum value of the electron-phonon coupling constant would be equal to $3$.} 
\label{f13}
\end{figure*}
%

%
\begin{figure*}[ht]
\includegraphics*[scale=0.6]{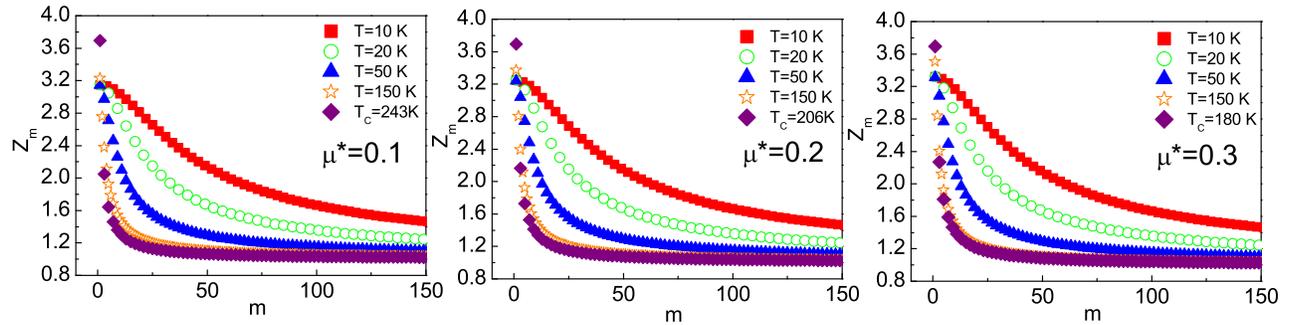}
\caption{The wave function renormalization factor on the imaginary axis for the selected values of the temperature and the Coulomb pseudopotential.} 
\label{f14}
\end{figure*}
%

%
\begin{figure}[ht]
\includegraphics*[width=\columnwidth]{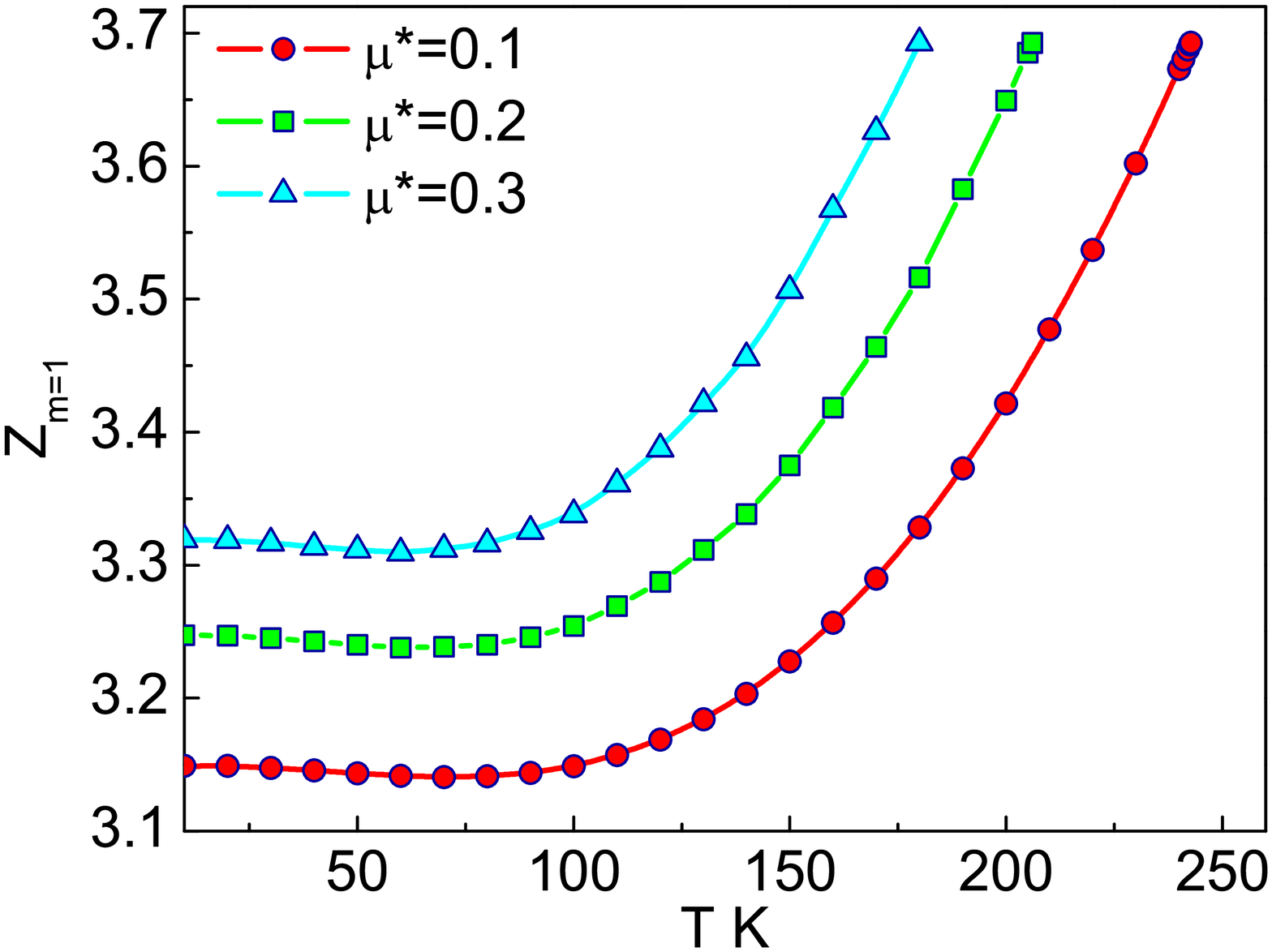}
\caption{The dependence of the wave function renormalization factor $Z_{m=1}$ on the temperature for the selected values of the Coulomb pseudopotential.
The functions presented in the figure can be parameterized by using the expression:
$Z_{m=1}\left(T,\mu^{\star}\right)=\left[Z_{m=1}\left(T_{C}\right)-Z_{m=1}\left(T_{0},\mu^{\star}\right)\right]\left(\frac{T}{T_{C}}\right)^{\beta}
+Z_{m=1}\left(T_{0},\mu^{\star}\right)$, where: $Z_{m=1}\left(T_{C}\right)=1+\lambda$, and $Z_{m=1}\left(T_{0},\mu^{\star}\right)=-1.297\left(\mu^{\star}\right)^{2}+1.369\mu^{\star}+3.025$.} 
\label{f15}
\end{figure}
%

In the presented paragraph, we have estimated the range of the values $R_{H}$ and $R_{C}$ for the hydrogenated compounds. For this purpose, the analytical formulas have been proposed, which have been derived based on the numerical results obtained for ${\rm CaH_{6}}$. In particular, they take the following form:

\begin{equation}
\label{r14}
\frac{R_{H}}{\left[R_{H}\right]_{{\rm BCS}}}=1-1.5\left(\frac{k_{B}T_{C}}{b\omega_{{\rm ln}}}\right)^{2}
\ln\left(\frac{b\omega_{{\rm \ln}}}{k_{B}T_{C}}\right),
\end{equation}
and
\begin{equation}
\label{r15}
\frac{R_{C}}{\left[R_{C}\right]_{{\rm BCS}}}=1+7\left(\frac{k_{B}T_{C}}{c\omega_{{\rm ln}}}\right)^{2}
\ln\left(\frac{c\omega_{{\rm \ln}}}{k_{B}T_{C}}\right),
\end{equation}
where: $b=0.345$ and $c=0.388$. The numerical and the analytical results have been presented together in \fig{f12}. The presented data demonstrate the correctness of the expressions \eq{r14} and \eq{r15}.

In the next step, the results have been generalized taking into account the Eliashberg function \eq{r6}. Assuming reasonable ranges of the values for the input parameters ($\lambda\in\left<1, 3\right>$ and $\mu^{\star}\in\left<0.1, 0.3\right>$), it has been found that the lowest value of the parameter $R_{H}$ is equal to $0.122$, whereas the highest value of $R_{C}$ equals $3.30$ (see also \fig{f13}). From the physical point of view, the presented estimations mean that the values of $R_{H}$ and $R_{C}$ obtained for the ${\rm CaH_{6}}$ compound determine the maximum derogation of the expectations of the BCS theory in the group of the hydrogen-rich compounds.
 
%
\section{THE ELECTRON EFFECTIVE MASS IN THE $\rm CaH_{6}$ COMPOUND}
%

The strong electron-phonon interaction, responsible for the formation of the superconducting state in the $\rm CaH_{6}$ compound, contributes to the significant renormalization of the electron mass. In the framework of Eliashberg formalism this effect has been determined by the wave function renormalization factor. 

\fig{f14} presents the form of the function $Z_{m}$ on the imaginary axis for the selected values of the temperature and the Coulomb pseudopotential. It can be noticed that with $m$ increasing, the values of $Z_{m}$ strongly decrease and then become saturated. The growth of the temperature causes successively quicker saturation of the wave function renormalization factor, such that for $T=T_{C}$ only a few dozens of the initial values make the significant contribution to the Eliashberg equation.

The influence of the Coulomb pseudopotential on the wave function renormalization factor can be investigated in the most convenient way by analyzing the behavior of $Z_{m=1}$. Based on \fig{f15}, it can be easily observed that with the increase of $\mu^{\star}$, the value of the wave function renormalization factor increases only slightly.

In the Eliashberg formalism, the quantity $Z_{m=1}$ plays a important role, because it determines the effective electron mass with a good approximation: $m^{\star}_{e}\simeq Z_{m=1}m_{e}$, where $m_{e}$ denotes the electron band mass. Analyzing the courses presented in \fig{f15}, it can be concluded that the effective mass of the electron is high in the entire area of the existence of the superconducting state. The parameter $m^{\star}_{e}$ only slightly increases with the growth of the temperature and reaches the maximum at the critical temperature. If $T=T_{C}$, the parameter $m^{\star}_{e}$ does not depend on the assumed value of the Coulomb pseudopotential and can be calculated analytically: $m^{\star}_{e}=\left(1+\lambda\right)m_{e}$. The obtained result is $m^{\star}_{e}= 3.69 m_{e}$. An identical result has been also obtained by the numerical meanings, which proves the high accuracy of the computer calculations.

The exact value of the parameter $m^{\star}_{e}$ should be estimated on the basis of the formula: 
$m^{\star}_{e}={\rm Re}\left[Z\left(0\right)\right]m_{e}$, where $Z\left(0\right)$ denotes the value of the wave function renormalization factor on the real axis for $\omega=0$. On the basis of Eqs. \eq{r1} and \eq{r2}, it has been found that the exact maximum effective mass equals $3.74 m_{e}$.

%
\section{SUMMARY}
%

In the paper, all the relevant thermodynamic parameters of the superconducting state in the ${\rm CaH_{6}}$ compound have been determined ($p=150$ GPa). 

It has been found that for the range of the Coulomb pseudopotential from $0.1$ to $0.3$, the critical temperature changes from $243$ K to $180$ K. 

Then, the values of the order parameter have been estimated. We have shown that the dimensionless ratio 
$R_{\Delta}$ very significantly differs from the prediction of the classical BCS theory. In particular, $R_{\Delta}\in\left<5.42, 5.02\right>$.

Also, the parameters associated with the thermodynamic critical field and the specific heat cannot be correctly estimated in the framework of the BCS model: $R_{H}\in\left<0.122,0.125\right>$ and $R_{C}\in\left<3.30,3.18\right>$.

The results obtained for ${\rm CaH_{6}}$ compound have been then generalized in such a way that it would be possible to estimate the values of the thermodynamic parameters in the group of the hydrogen-rich compounds. 

It has been shown that for the reasonable values of the input parameters, the maximum value of the critical temperature is equal to $764$ K. The obtained data correlates well with the maximum value of $T_{C}$, estimated for metallic atomic hydrogen ($p=2$ TPa) \cite{Maksimov}, \cite{Szczesniak4}. Thus, from the physical point of view, the achieved result means that in the group of the hydrogenated compounds the systems may exist which are characterized by the critical temperature comparable to the room temperature at the relatively low pressure.

According to the remaining thermodynamic parameters ($R_{\Delta}$, $R_{H}$, and $R_{C}$), it has been stated that their values in the group of the hydrogen-rich compounds should not deviate from the predictions of the BCS theory more than the analogous parameters in ${\rm CaH_{6}}$.

\section{Acknowledgments}
\label{acknowledgments}

The authors would like to thank Prof. K. Dzili{\'n}ski for creating excellent working conditions.

The numerical calculations for ${\rm CaH_{6}}$ compound have been based on the Eliashberg function sent to us by: Prof. Yanming Ma and Prof. Hui Wang to whom we are very thankful.

Additionally, we are grateful to the Cz{\c{e}}stochowa University of Technology - MSK CzestMAN for granting access to the computing infrastructure built in the project No. POIG.02.03.00-00-028/08 "PLATON - Science Services Platform".
%
\appendix
%
%
\section{The value of the critical temperature for the metallic hydrogen}

In the Appendix we have calculated the critical temperature for metallic hydrogen by using the formula \eq{r5}. We have considered the selected values of the pressure. The results from \tab{t2} have proved the validly of the presented analytical approach.
%
\begin{table*}[!h]
\caption{\label{t2} The critical temperature for the metallic hydrogen.}
\begin{ruledtabular}
\begin{tabular}{cccccccc}
 $p$ GPa & $\lambda$ & $\omega_{\rm ln}$ meV & $\sqrt{\omega_{2}}$ meV & $\mu^{\star}$ & $T_{C}$ K (Eliashberg) & $T_{C}$ K (Eq. (5)) & Ref. \\
\hline
 $347$  & $0.93$ & $141.99$  & $199.01$ & $0.08-0.15$ & $120-90$   & $113-85$  & \cite{Szczesniak2}   \\
 $428$  & $1.20$ & $141.95$  & $207.48$ & $0.08-0.15$ & $179-141$  & $176-140$ & \cite{Szczesniak25}  \\
 $480$  & $2.21$ & $153.77$  & $195.99$ & $0.1-0.3$   & $-$        & $304-255$ & \cite{Yan}           \\
 $539$  & $2.04$ & $166.71$  & $208.00$ & $0.1-0.3$   & $360$      & $307-252$ & \cite{Szczesniak5}   \\
 $608$  & $1.91$ & $174.25$  & $217.84$ & $0.1-0.3$   & $-$        & $302-242$ & \cite{Yan}           \\
 $802$  & $1.70$ & $184.86$  & $237.82$ & $0.1-0.3$   & $332- 259$ & $285-218$ & \cite{Szczesniak24}  \\   
 $2000$ & $7.32$ & $89.20$   & $161.07$ & $0.1-0.5$   & $631-413$  & $719-506$ & \cite{Szczesniak4}   \\\hline                                                                                             
\end{tabular}
\end{ruledtabular}
\end{table*}


%
%

\begin{thebibliography}{99}
%
\bibitem{Ashcroft}
N.W. Ashcroft, Phys. Rev. Lett. {\bf 92}, 187002 (2004).
\bibitem{Tse}
J.S. Tse, Y. Yao, K. Tanaka, Phys. Rev. Lett. {\bf 98}, 117004 (2007).
\bibitem{Gao}
G. Gao, A.R. Oganov, A. Bergara, M. Martinez-Canales, T. Cui, T. Iitaka, Y. Ma, G. Zou, Phys. Rev. Lett. {\bf 101}, 107002 (2008).
\bibitem{Canales}
M. Martinez-Canales, A.R. Oganov, Y. Ma, Y. Yan, A.O. Lyakhov, A. Bergara, Phys. Rev. Lett. {\bf 102}, 87005 (2009).
\bibitem{Gao1}
G. Gao, A.R. Oganov, P. Li, Z. Li, H. Wang, T. Cui, Y. Ma, A. Bergara, A.O. Lyakhov, T. Iitaka, G. Zou, 
Proc. Nat. Acad. Sci. USA {\bf 107}, 1317 (2010).
\bibitem{Chen}
X.J. Chen, V.V. Struzhkin, Y. Song, A.F. Goncharov, M. Ahart, Z. Liu, H. Mao, R.J. Hemley, 
Proc. Nat. Acad. Sci. USA {\bf 105}, 20 (2008).
\bibitem{Eremets}
M.I. Eremets, I.A. Trojan, S.A. Medvedev, J.S. Tse, Y. Yao, Science {\bf 319}, 1506 (2008).
\bibitem{Stadele}
M. Stadele, R.M. Martin, Phys. Rev. Lett. {\bf 84}, 6070 (2000).
\bibitem{Cudazzo01}
P. Cudazzo, G. Profeta, A. Sanna, A. Floris, A. Continenza, S. Massidda, E.K.U. Gross, Phys. Rev. Lett. {\bf 100}, 257001 (2008).
\bibitem{Cudazzo02}
P. Cudazzo, G. Profeta, A. Sanna, A. Floris, A. Continenza, S. Massidda, E.K.U. Gross, Phys. Rev. B {\bf 81}, 134505 (2010).
\bibitem{Cudazzo03}
P. Cudazzo, G. Profeta, A. Sanna, A. Floris, A. Continenza, S. Massidda, E.K.U. Gross, Phys. Rev. B {\bf 81}, 134506 (2010).
\bibitem{Zhang}
L. Zhang, Y. Niu, Q. Li, T. Cui, Y. Wang, Y. Ma, Z. He, G. Zou, Solid State Commun. {\bf 141}, 610 (2007).
\bibitem{Szczesniak1}
R. Szcz{\c{e}}{\'s}niak, M.W. Jarosik, Physica B {\bf 406}, 3493 (2011).
\bibitem{Szczesniak2}
R. Szcz{\c{e}}{\'s}niak, M.W. Jarosik, Physica B {\bf 406}, 2235 (2011).
\bibitem{Szczesniak3}
R. Szcz{\c{e}}{\'s}niak, E.A. Drzazga, accepted in: Solid State Sciences; 
preprint: arXiv:1209.5849 (2012).
\bibitem{BCS1}
J. Bardeen, L.N. Cooper, J.R. Schrieffer, Phys. Rev. {\bf 106}, 162 (1957).
\bibitem{BCS2} 
J. Bardeen, L.N. Cooper, J.R. Schrieffer, Phys. Rev. {\bf 108}, 1175 (1957).
\bibitem{Yan}
Y. Yan, J. Gong, Y. Liu, Phys. Lett. A 375, 1264 (2011).
\bibitem{Maksimov}
E.G. Maksimov, D.Y. Savrasov, Solid State Commun. {\bf 119}, 569 (2001).
\bibitem{McMahon}
J.M. McMahon, D.M. Ceperley, Phys. Rev. B {\bf 84}, 144515 (2011).
\bibitem{McMahon1}
J.M. McMahon, D.M. Ceperley, Phys. Rev. B {\bf 85}, 219902(E) (2012).
\bibitem{Liu}
H. Liu, H. Wang, Y. Ma, J. Phys. Chem. C {\bf 116}, 9221 (2012).
\bibitem{Szczesniak4}
R. Szcz{\c{e}}{\'s}niak, M.W. Jarosik, Solid State Commun. {\bf 149}, 2053 (2009).
\bibitem{Szczesniak5}
R. Szcz{\c{e}}{\'s}niak, D. Szcz{\c{e}}{\'s}niak, E.A. Drzazga, Solid State Commun. {\bf 152}, 2023 (2012).
\bibitem{Szczesniak6}
R. Szcz{\c{e}}{\'s}niak, PLoS ONE 7 (4), art. no. e31873 (2012);  
preprint: arXiv:1105.5525 (2011) and arXiv:1110.3404 (2012).
\bibitem{Szczesniak7}
R. Szcz{\c{e}}{\'s}niak, A.P. Durajski, preprint: arXiv:1206.5531 (2012).
\bibitem{Li}
Y. Li, G. Gao, Y. Xie, Y. Ma, T. Cui, G. Zou, Proc. Nat. Acad. Sci. USA {\bf 107}, 15708 (2010).
\bibitem{Jin}
X. Jin, X. Meng, Z. He, Y. Ma, B. Liu, T. Cui, G. Zou, H. Mao, Proc. Nat. Acad. Sci. USA {\bf 107}, 9969 (2010).
\bibitem{Kazutaka}
A. Kazutaka, N.W. Ashcroft, Phys. Rev. B {\bf 84}, 104118 (2011).
\bibitem{Kim}
D.Y. Kim, R.H. Scheicher, C.J. Pickard, R.J. Needs, R. Ahuja, Phys. Rev. Lett. {\bf 107}, 117002 (2011).
\bibitem{Deemyad}
S. Deemyad, J.S. Schilling, Phys. Rev. Lett. 91, 167001 (2003).
\bibitem{Yabuuchi}
T. Yabuuchi, T. Matsuoka, Y. Nakamoto, K. Shimizu, J. Phys. Soc. Jpn. 75, 083703 (2006).
\bibitem{Sakata}
M. Sakata, Y. Nakamoto, K. Shimizu, T. Matsuoka, Y. Ohishi, Phys. Rev. B {\bf 83}, 220512(R) (2011).
\bibitem{Andersson}
M. Andersson, Phys. Rev. B {\bf 84}, 216501 (2011).
\bibitem{Szczesniak8}
R. Szcz{\c{e}}{\'s}niak, M.W. Jarosik, D. Szcz{\c{e}}{\'s}niak, Physica B {\bf 405}, 4897 (2010).
\bibitem{Szczesniak9}
R. Szcz{\c{e}}{\'s}niak, A.P. Durajski, Physica C {\bf 472}, 15 (2012).
\bibitem{Szczesniak10}
R. Szcz{\c{e}}{\'s}niak, A.P. Durajski, Journal of Superconductivity and Novel Magnetism {\bf 25}, 399 (2012).
\bibitem{Szczesniak11}
R. Szcz{\c{e}}{\'s}niak, A.P. Durajski, M.W. Jarosik, Mod. Phys. Lett. B {\bf 26}, 1250050 (2012). 
\bibitem{Szczesniak12}
R. Szcz{\c{e}}{\'s}niak, A.P. Durajski, Solid State Commun. {\bf 152}, 1018  (2012).
\bibitem{Szczesniak13}
R. Szcz{\c{e}}{\'s}niak, D. Szcz{\c{e}}{\'s}niak, Physica Status Solidi B, {\bf 249}, 2194 (2012).
\bibitem{Wang}
H. Wang, J.S. Tse, K. Tanaka, T. Iitaka, Y. Ma, Proc. Nat. Acad. Sci. USA {\bf 109} (17), 6463 (2012).
\bibitem{Marsiglio}
F. Marsiglio, M. Schossmann, J.P. Carbotte, Phys. Rev. B {\bf 37}, 4965 (1988).
\bibitem{Eliashberg1}
G.M. Eliashberg, Soviet. Phys. JETP {\bf 11}, 696 (1960).
\bibitem{Eliashberg2}
P.B. Allen, B. Mitrovi{\'c}, in: Solid State Physics: Advances in Research and Applications,
edited by H. Ehrenreich, F. Seitz, D. Turnbull, (Academic, New York, 1982), Vol 37, p. 1.  
\bibitem{Eliashberg3}
J.P. Carbotte, Rev. Mod. Phys. {\bf 62}, 1027 (1990).
\bibitem{Eliashberg4}
J.P. Carbotte, F. Marsiglio, in: The Physics of Superconductors, 
edited by K.H. Bennemann, J.B. Ketterson, (Springer, Berlin, 2003), Vol 1, p. 223.
\bibitem{Szczesniak14}
A.P. Durajski, R. Szcz{\c{e}}{\'s}niak, M.W. Jarosik, Phase Transitions, {\bf 85}, 727 (2012).
\bibitem{Szczesniak15}
R. Szcz{\c{e}}{\'s}niak, Solid State Commun. {\bf 145}, 137 (2008).
\bibitem{Szczesniak16}
R. Szcz{\c{e}}{\'s}niak, M. Mierzejewski, J Zieli{\'n}ski, Physica C {\bf 355}, 126 (2001).
\bibitem{Szczesniak17}
R. Szcz{\c{e}}{\'s}niak, D. Szcz{\c{e}}{\'s}niak, Solid State Commun. {\bf 152}, 779 (2012).
\bibitem{AllenDynes}
P.B. Allen, R.C. Dynes, Phys. Rev. B {\bf 12}, 905 (1975). 
\bibitem{McMillan}
W.L. McMillan, Phys. Rev. {\bf 167}, 331 (1968).
\bibitem{Szczesniak18}
R. Szcz{\c{e}}{\'s}niak, A.P. Durajski, Solid State Commun. {\bf 153}, 26 (2013).
\bibitem{Durajski}
A.P. Durajski, Physica C {\bf 485}, 145 (2013).
\bibitem{Szczesniak19}
R. Szcz{\c{e}}{\'s}niak, A.P. Durajski, J. Phys. Chem. Solids {\bf 74}, 641 (2013).
\bibitem{Szczesniak20}
R. Szcz{\c{e}}{\'s}niak, E.A. Drzazga, A.M. Duda, preprint: arXiv:1303.0500 (2013).
\bibitem{Szczesniak21}
R. Szcz{\c{e}}{\'s}niak, A.P. Durajski, D. Szcz{\c{e}}{\'s}niak, preprint: arXiv:1212.2356 (2013).
\bibitem{Szczesniak22}
R. Szcz{\c{e}}{\'s}niak, A.P. Durajski, preprint: arXiv:1302.3050 (2013).
\bibitem{Szczesniak23}
R. Szcz{\c{e}}{\'s}niak, D. Szcz{\c{e}}{\'s}niak, K.M. Huras, preprint: arXiv:1303.1223 (2013).
\bibitem{Varelogiannis}
G. Varelogiannis, Z. Phys. B {\bf 104}, 411 (1997). 
\bibitem{Bardeen}
J. Bardeen, M. Stephen, Phys. Rev. {\bf 136}, A1485 (1964).
\bibitem{Szczesniak24}
Private information: A.P. Durajski.
\bibitem{Szczesniak25}
R. Szcz{\c{e}}{\'s}niak, M.W. Jarosik, Acta Phys. Pol. A {\bf 121}, 841 (2012). 
%
\end{thebibliography}
\end{document}